\documentclass[preprint, 5p, amsmath, fleqn]{elsarticle}

\usepackage{hyperref}
\usepackage{mathtools}
\usepackage{float}

\journal{Physics Letters B}

\biboptions{numbers,sort&compress}

\def\p{\\[0.3cm]}
\def\<{\langle} 			
\def\>{\rangle} 
\def\nn{\nonumber} 		
\newcommand{\fr}[2]{{\textstyle\frac{#1}{#2}}}
\newcommand{\ffr}[2]{\ensuremath{\frac{\displaystyle #1}{\displaystyle #2}}}

\begin{document}

\begin{frontmatter}

\title{Charge symmetry violation in the doubly charmed cascade masses}

\author{K. K. Cushman\fnref{myfootnote}}
\fntext[myfootnote]{Currently at Yale University Department of Physics}

\author{A. W. Thomas}

\author{R. D. Young}

\address{CSSM \& CoEPP, Department of Physics, University of Adelaide, Adelaide SA 5005, Australia}

\begin{abstract}
 We investigate the electromagnetic contribution to the charge symmetry breaking in the $\Xi_{cc}$ baryon masses using a subtracted dispersion relation based on the Cottingham formula, following the formalism developed in an analysis of the octet baryon mass differences. In the absence of experimental information on the structure of charmed baryons, we use parameters for the electromagnetic structure of the $\Xi_{cc}$ and the difference in its charge states obtained from lattice QCD and estimates of SU(4) symmetry breaking. We report a conservative estimate for the mass splitting of the doubly charmed cascades to be 8 $\pm$ 9 MeV. While a smaller isospin splitting is compatible with this result, surprisingly it does not preclude the large splitting reported by the SELEX Collaboration. We identify those quantities which could be determined from lattice QCD and which would reduce the quoted theoretical uncertainty.
\end{abstract}


\end{frontmatter}


\section{\label{sec:intro}Introduction}
The total charge symmetry violation (CSV) in the masses of a baryon multiplet is a result of symmetry breaking arising from quark mass differences and electromagnetic forces. The inequality of the quark masses ($m_u \neq m_d$) is referred to as the strong component, while the energy difference arising from the electromagnetic interaction is referred to as the electromagnetic contribution. Thus, for small mass differences and weak electromagnetic coupling, the charge symmetry breaking for a baryon $B$ is given by the sum $\delta M_B = \delta M_B^{\rm strong} + \delta M_B^{\gamma}$. Experimentally, the approximate mass splitting of the nucleon is $ M_p - M_n \, \approx \, -1.3$ MeV~\cite{Patrignani:2016xqp}. More exotic baryon states, such as the singly charmed sigmas have an isospin mass splitting opposite to that of most particles, with $M_{\Sigma_c^{++}} - M_{\Sigma_c^{+}} \approx   1$ MeV \cite{Patrignani:2016xqp}.

It is notable that nearly all of the baryon isospin pairs have mass differences ranging from about 7 MeV to a couple of MeV~\cite{Patrignani:2016xqp}. However, in 2002 the SELEX collaboration at Fermilab reported the observation of two families of doubly charmed cascades, the $\Xi_{cc}^+$(3443) and $\Xi_{cc}^{++}$(3460) forming one isospin doublet, and the $\Xi_{cc}^+$(3520) and $\Xi_{cc}^{++}$(3541) forming another \cite{hep-ex/0208014,hep-ex/0212029,selex3}. Thus the isospin mass splittings found by SELEX were reported to be 17 MeV and 21 MeV, respectively~\cite{selex3}. These mass splittings are much larger than those of any other hadronic isospin pair. To further complicate this puzzle, the LHCb Collaboration have reported the observation of a doubly-charged $\Xi_{cc}^{++}$ with a mass $3621\,\mathrm{MeV}$. While we await further observations of the $\Xi_{cc}$ baryons to resolve the spectrum, it is of interest to investigate the degree of isospin symmetry breaking in these states. 

The possibility of large mass splitting as observed by SELEX is of theoretical interest because it may indicate something extraordinary about the quark structure of heavy and doubly heavy baryons~\cite{Brodsky:2011zs, Can:2013tna}.  Estimates of the strong contribution to $M_p - M_n$ report this to be  $\sim -2$--3 MeV~\cite{Beane:2006fk, Horsley:2012fw,Shanahan:2012wa,Borsanyi:2014jba,Horsley:2015eaa}.
The corresponding calculations of the doubly-strange cascade baryons suggest a strong contribution to the splitting $M_{\Xi^0}-M_{\Xi^-}$ on the order of $\sim -5$--$6\,\mathrm{MeV}$. 
The doubly-charmed baryon would be anticipated to be of similar magnitude and, importantly, of the same sign.
Thus, in order for CSV to yield a large and negative mass splitting for the  doubly-charmed cascades, $M_{\Xi_{cc}^{++}}-M_{\Xi_{cc}^{+}}$, the electromagnetic self energy would have to be greater than the total mass splitting and thus as large as $\sim 20\,\mathrm{MeV}$ in magnitude.

In this work, we provide an estimate of the electromagnetic charge symmetry breaking $\delta M_{\Xi_{cc}^{++}}^{\gamma} - \delta M_{\Xi_{cc}^{+}}^{\gamma}$ based on the Cottingham sum rule \cite{Cottingham:1963zz,Gasser:1974wd,Collins:1978hi}.
Since little experimental information is available for the structure of baryons beyond the octet, we use lattice results of the electromagnetic structure and estimates of SU(4) symmetry breaking to guide the analysis in its application to the exotic $\Xi_{cc}$ systems. 

\section{\label{sec:EM_self_energy}Electromagnetic Self-Energy}
A dispersion relation analysis of the electromagnetic self energy of a baryon $B$ may be written 
as a sum of four contributions~\cite{Gasser:1974wd,WalkerLoud:2012bg,Gasser:2015dwa}
\begin{eqnarray}
\delta M_B^{\gamma} = \delta M_B^{\rm el} + \delta M_B^{\rm inel} + \delta M_B^{\rm sub} + \delta M_B^{\rm ct} \, . \label{contributions}
\end{eqnarray}
In the following sections we explore each term, adopting the formalism of Ref.~\cite{Erben:2014hza} which provides a minor revision of the analysis presented by Walker-Loud {\it et al.}~\cite{WalkerLoud:2012bg}. 
%

\subsection{Elastic Contribution}
The elastic contribution to the electromagnetic self energy is given by 
\begin{eqnarray}\label{elastic}
&\delta M_B^{\rm el} &= \ffr{\alpha}{\pi} \int_0^{\Lambda_0} dQ \Bigg[\ffr{3}{2}G_M^2 \ffr{\sqrt{\tau_{\rm el}}}{\tau_{\rm el} + 1} \nn\\[0.1cm] 
&&\,\,\,\,+(G_E^2- 2\tau_{\rm el} G_M^2)\ffr{(1+\tau_{\rm el})^{3/2} - \tau_{\rm el}^{3/2} - \fr{3}{2}\sqrt{\tau_{\rm el}}}{\tau_{\rm el} + 1} \Bigg] \, ,
\nn \\ 
\end{eqnarray}
where $\Lambda_0$ is a renormalization scale, chosen to lie in a range above which elastic contributions are negligible. Numerically, we set the renormalization constant $\Lambda_0^2 = 20$ GeV$^2$ and calculate uncertainties by allowing variation over the range 10 $< \Lambda_0^2 <$ 30 GeV$^2$. We find that the results are insensitive to the choice of $\Lambda_0$ in this range. The kinematic factor $\tau_{\rm el} = Q^2/(4M_B^2)$, with $M_B$ the baryon mass. We allow for a $\Xi_{cc}$ mass in the range 3.45 $< M_{\Xi_{cc}} <$ 3.65 GeV in light of the LHCb collaboration recently reporting the observation of $\Xi_{cc}^{++}$ with a mass of 3621.40(78) MeV~\cite{Aaij:2017ueg}. For the moment it is unclear how this relates to the states observed by SELEX. $G_E$ and $G_M$ are the electric and magnetic elastic form factors for the baryon. There is no experimental data on the charmed cascade form factors, so we are guided by the empirical fits of the nucleon form factors which take a dipole form:
\begin{align} 
G_{E,M}^B(Q^2) &= \ffr{G_{E,M}^B(0)}{\big(1+\fr{Q^2\<r_{E,M}^2\>^B}{12}\big)^2} \, .
\end{align} 
The factors $\<r_{E,M}^2\>^B$ are the electric and magnetic charge radii of the baryon $B$, $G_E^B(0)$ is the electric charge in units of $e$, and $G_M^B(0)$ is proportional to the baryon magnetic moment with $\mu_B = \fr{e}{2M_B}G_M(0)$. Since experimental data about form factors of the $\Xi_{cc}$ are unavailable, we rely on the lattice computation of these electromagnetic form factors reported in Ref.~\cite{Can:2013tna}. The baryon form factors can be formulated as a sum of individual quark-sector contributions, where the total form factors for $\Xi_{cc}^{++/+}$ are given as
\begin{align}
G_{E}^{\Xi_{cc}^{++}} &= 2\mathcal{Q}_c G_E^{\Xi_{cc}^{++},c}(Q^2) + \mathcal{Q}_{u} G_E^{\Xi_{cc}^{++}, u}(Q^2),\\[0.3cm]
G_{M}^{\Xi_{cc}^{++}} &=  G_M^{\Xi_{cc}^{++},c}(Q^2) + G_M^{\Xi_{cc}^{++}, u}(Q^2), \\[0.3cm]
G_{E}^{\Xi_{cc}^{+}} &= 2\mathcal{Q}_c G_E^{\Xi_{cc}^{+},c}(Q^2) + \mathcal{Q}_{d} G_E^{\Xi_{cc}^{+}, d}(Q^2),\\[0.3cm]
G_{M}^{\Xi_{cc}^{+}} &=  G_M^{\Xi_{cc}^{+},c}(Q^2) + G_M^{\Xi_{cc}^{+}, d}(Q^2).
\end{align}
Note that $\mathcal{Q}_{u,d,c}$ are the charges of the respective quarks. 
The numerical values of the charge radii used for the elastic integral were obtained from Ref.~\cite{Can:2013tna}, and the magnetic moments,
$\mu_{\Xi_{cc}^{++}}$ and $\mu_{\Xi_{cc}^+}$ were determined from an average of results based on quark models, chiral effective theories, and lattice calculations~\cite{Can:2013tna, Barik:1984tq, JuliaDiaz:2004vh, Kumar:2005ei, hep-ph/0602193, hep-ph/0610030, 1003.4338, 1209.2900,1707.02765}, taking uncertainties to allow for variation over all of the values obtained in the literature. The charge radii and magnetic moments used are given in Table~\ref{elastic_params}. 

\begin{figure}[H]
\begin{center}
\caption{(Color online) Elastic integrand as given in Eq.~\ref{elastic} for the $\Xi_{cc}^{++}$ and $\Xi_{cc}^{+}$ as well as for the proton and neutron. }
\includegraphics[width = 0.95\linewidth]{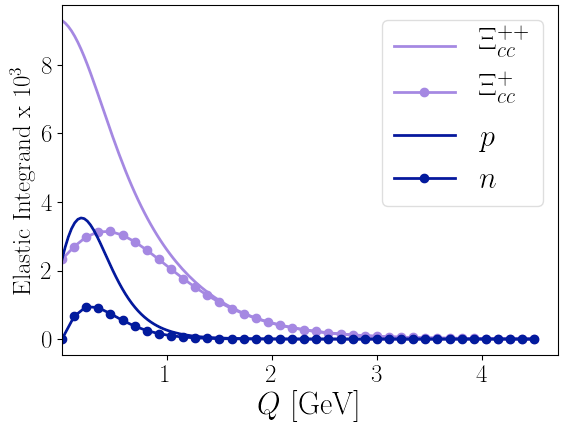}
\label{elastic_compare}
\end{center}
\end{figure}

In Fig.~\ref{elastic_compare} we illustrate the elastic integrands (Eq.~\ref{elastic}) for the $\Xi_{cc}^{++}$ and $\Xi_{cc}^+$.  These are compared with the elastic integrands for the proton and neutron computed using form factors determined by the Kelly parametrization~\cite{Kelly:2004hm}, as used in the dispersive analysis for the octet baryons in Ref.~\cite{Erben:2014hza}. In the case of the octet baryon isospin pairs, $N$, $\Sigma$ and $\Xi$, the elastic self-energy was found to provide the largest contribution to the electromagnetic self-energy difference between charge states. The difference in the elastic self-energy between the proton and neutron is represented by the area between the dark blue curves in Fig.~\ref{elastic_compare}. The difference between the elastic self-energies of the the doubly charmed cascades (area between the lavender curves), determined using the parameters in Table~\ref{elastic_params}, clearly yields a greater contribution to the CSV between the two charge states. 
The result for the elastic contribution to the mass splitting of the $\Xi_{cc}$ is found in Table \ref{results_table}. 
\begin{table}[H]
\begin{center}
\caption{Electric and magnetic parameters for $\Xi_{cc}^{++}$ and $\Xi_{cc}^+$. 
Charge radii given in fm$^2$, and magnetic moments given in nuclear magnetons, determined from Refs.~\cite{Can:2013tna, Barik:1984tq, JuliaDiaz:2004vh, Kumar:2005ei, hep-ph/0602193, hep-ph/0610030, 1003.4338, 1209.2900,1707.02765}.}
\begin{tabular}{|c|c|}
\hline 
Parameter & Value\\
\hline 
$\<r^2_{E, c}\>$ & 0.095(9)\\
$\<r^2_{E, u/d}\>$ & 0.410(46)\\
\hline 
$\<r^2_{M, c}\>$ & 0.089(11)\\
$\<r^2_{M, u/d}\>$ & 0.612(115)\\
\hline
$\mu_{\Xi_{cc}^{++}}$ & 0.75(30)\\
\hline 
$\mu_{\Xi_{cc}^+}$ & -0.10(10)\\
\hline 
\end{tabular}
\label{elastic_params} 
\end{center}
\end{table}
%
%
\subsection{Inelastic Contribution}
Following Refs.~\cite{Erben:2014hza, WalkerLoud:2012bg}, the inelastic contribution to the electromagnetic self-energy is given as
\begin{align}
\delta M_B^{\rm inel} &= \int_{W_0^2}^{\infty} dW^2\Omega_B^{\rm inel}(W^2) \, ,
\label{resonances}
\end{align}
where $W_0$ is the threshold for excited states. $\Omega_B^{\rm inel}(W^2)$ is the inelastic contribution to the self-energy associated with 
each intermediate hadronic state of invariant mass squared $W^2$. As in Ref.~\cite{Erben:2014hza}, $\Omega_B^{\rm inel}(W^2)$ is given by
\begin{eqnarray}
\label{Omega}
&&\hspace{-0.5cm}\Omega_B^{\rm inel}(W^2) = \nn\p
&&\,\,\,\ffr{\alpha}{\pi}\int_0^{\Lambda_0} dQ \Bigg\{\ffr{3F_1^B(W^2, Q^2)}{4M_B^2}\,\,\ffr{2\tau^{3/2}-2\tau\sqrt{1+\tau}+\sqrt{\tau}}{\tau} \nn\p 
&&\,\,\, + \ffr{F_2^B(W^2,Q^2)}{(Q^2+W^2-M_B^2)}\Big[(1+\tau)^{3/2}-\tau^{3/2} -\fr{3}{2}\sqrt{\tau}\Big]\Bigg\} \, ,
\nn\\
\end{eqnarray}
where $\tau$ is the kinematic factor given by, $\tau = (W^2+Q^2-M_B^2)^2/(4M_B^2Q^2)$. $F_1^B$ and $F_2^B$ are the inelastic structure functions for the baryon. While we do not have empirical constraints on the structure functions for the charmed cascades, we can at least estimate the results of the integral~\ref{Omega} for these exotic baryons by calculating an approximate structure function for the $\Xi_{cc}^{++/+}$ based on the quark momentum fractions, $\langle x\rangle$, that we expect each valence quark to carry. That is, we use the structure functions

\begin{align}
F_1^B(x, Q^2)&= \ffr{1}{2}\sum_q \mathcal{Q}_q^2 f^B_q(x, Q^2)\\[0.2cm]
F_2^B(x, Q^2) &= x\sum_q \mathcal{Q}_q^2 f^B_q(x, Q^2) \, ,
\end{align}
where we sum over the relevant quark flavors, in this case $u,d,s,c$. $\mathcal{Q}_q$ is the charge of quark $q$ and $f^B_q(x, Q^2)$ is the parton density function of quark $q$ in baryon $B$ at momentum fraction $x$ and resolution determined by $Q^2$. Taking the nucleon as an example and following Ref.~\cite{Erben:2014hza}, we assume charge symmetry~\cite{Londergan:2009kj}
 in the nucleon. In the case of the nucleon this means $f^p_u = f^n_d \equiv f^N_u$ and $f^p_d = f^n_u \equiv f^N_d$, where the doubly or singly represented quark densities are assumed to be the same in each nucleon. Thus, 
\begin{align}
f^N_u &= 2\cdot\ffr{9}{15}(4F_1^p - F_1^n) \\[0.1cm]
&= \ffr{1}{x}\cdot \ffr{9}{15}(4F_2^p - F_2^n)
\end{align}
and
\begin{align}
f^N_d &= 2\cdot \ffr{9}{15}(4F_1^n - F_1^p) \\[0.1cm]
&= \ffr{1}{x}\cdot \ffr{9}{15}(4F_2^n - F_2^p) \, .
\end{align} 
Based on the ratios of the constituent quark masses, $M_{q}$, of the valence quarks in the $\Xi_{cc}^{++/+}$ and the nucleon, we estimate the ratio of the quark momentum fractions to be 
\begin{align}
\ffr{\<x\>^{\Xi_{cc}}_c}{\<x\>^p_u} \,\,\,\approx \,\,\,\ffr{M_c/M_{\Xi_{cc}}}{M_u/M_p},\p 
\ffr{\<x\>^{\Xi_{cc}}_l}{\<x\>^p_d}\,\,\,\approx\,\,\, \ffr{M_l/M_{\Xi_{cc}}}{M_d/M_p} ,
\end{align}
where $l$ represents the light quark in the $\Xi_{cc}$. Following Ref.~\cite{Erben:2014hza}, we use these ratios to scale the nucleon structure functions appearing in Eq.~\ref{Omega}. It follows from the linearity of Eq. \ref{Omega} in $F_1$ and $F_2$ that 
\begin{align}\label{inelastic_before}
\hspace{-1cm}\delta M_{\Xi_{cc}^{++}}^{\rm inel} - \delta M_{\Xi_{cc}^{+}}^{\rm inel} &= \ffr{9}{15}(\mathcal{Q}_u^2 -\mathcal{Q}_d^2)\ffr{\<x\>^{\Xi_{cc}}_l}{\<x\>^p_d}(4\delta M_n^{\rm inel} - \delta M_p^{\rm inel}).
\end{align}
We take the nucleon inelastic results from Ref.~\cite{Erben:2014hza} to be $\delta M_p^{\rm inel} = 0.62(8)$ MeV and $\delta M_n^{\rm inel} =  0.53(7)$ MeV. Constituent quark masses for the proton are taken to be $M_u$ = 336 MeV, $M_d$ = 340 MeV. The electromagnetic structure of the $\Xi_{cc}$ appears to vary from that of the nucleon, i.e. smaller charge radii of the heavy $c$ quarks as determined in Refs.~\cite{Can:2013tna, Beane:2006fk}, so we use the constituent quark masses $M_c$ = 1486 MeV and $M_l$ = 385 MeV, as given in an investigation of heavy baryon spectroscopy based on a quark-diquark model~\cite{Gershtein:2000nx}. Thus we find,  
\begin{align}
\ffr{\<x\>^{\Xi_{cc}}_l}{\<x\>^p_d} &= 0.30 \, , 
\end{align} 
which yields a result for Eq. \ref{inelastic_before} of 0.10 MeV. 

The resonance structures of the $\Xi_{cc}^{++/+}$ would however be markedly different than that those of the nucleon. In particular, the reduction of the hyperfine interaction caused by the larger charm quark mass could reduce the splitting between the ground state and the spin-3/2 excited state by as much as a factor of 3. In  Ref.~\cite{Gershtein:2000nx} the spectrum of doubly heavy baryons is predicted using the quark-diquark model. The doubly charmed cascade is predicted to have resonances of the ground state $\Xi_{cc}^{++}$ and $\Xi_{cc}^+$ that are relatively closely spaced compared to the nucleon resonances. For example, the two lowest resonant excitations of the nucleon lie within  293 MeV ($\Delta$), and 500 MeV (Roper resonance) of the ground state nucleon, while the lowest two excitations of the $\Xi_{cc}$ are predicted to lie within 132 MeV and 224 MeV \cite{Gershtein:2000nx}. Experimentally, SELEX proposed possible observations of $\Xi_{cc}$ resonances lying within 320 MeV of each other \cite{hep-ex/0212029}. As the contribution from the inelastic term calculated in the present section is relatively small compared to the contributions from the other terms it seems reasonable to seek an order of magnitude estimate of the inelastic contribution by scaling the result of Eq.~\ref{inelastic_before} to account for the difference in the spectrum of the $\Xi_{cc}$ compared with the nucleon. To this end, we scale the result of Eq.~\ref{inelastic_before} by a factor of three since in a given range over $W^2$, the $\Xi_{cc}$ would have about three times more resonance structures contributing to the integral \ref{resonances}. Thus, we take
\begin{align}
\delta M_{\Xi_{cc}^{++}}^{\rm inel} - \delta M_{\Xi_{cc}^{+}}^{\rm inel} &\approx 3\times 0.10 {\rm \, MeV}. 
\end{align}
Based on this hypothesis we attach an uncertainty of 0.2 MeV to the earlier calculation, as summarized in Table~\ref{results_table}.

\subsection{Subtraction Terms}
Following Ref.~\cite{Erben:2014hza} and the analysis outlined in Ref.~\cite{WalkerLoud:2012bg}, we include a subtraction term given by 
\begin{align}
\delta M_B^{\rm sub} &= -\ffr{3\alpha}{16\pi M_B}\int_0^{\Lambda_0^2} dQ^2\, T_1^B(0,Q^2) \, ,
\end{align} 
where $\Lambda_0$ is the renormalization scale, as before, and the amplitude $T_1^B(0, Q^2)$ is a scalar function related to the Lorentz contracted Compton tensor, $T^{\mu}_{\mu}$. See Ref.~\cite{WalkerLoud:2012bg} for the decomposition of this tensor. The momentum dependence of $T_1^B$ can be summarized in a model independent fashion as 
\begin{align} 
\hspace{-0.8cm}T_1^B(0, Q^2) = 2[G_M^B(Q^2)]^2 - 2[F_D^B(Q^2)]^2 + T_1^{B, \rm inel}(Q^2) \, , 
\nn  \\ 
\end{align} 
where $G_M^B(Q^2)$ and $F_D^B(Q^2)$ are the magnetic and Dirac elastic form factors for the baryon $B$. This elastic portion of the integral can be readily computed using the elastic form factors computed for $\Xi_{cc}^{++}$ and $\Xi_{cc}^+$ from the elastic term. See Table \ref{results_table} for a summary of this result.

The remaining term in the subtraction component is an inelastic portion, which can be modeled based on knowledge of the low $Q^2$ and high $Q^2$ behavior of the function. Following Ref.~\cite{Erben:2014hza}, we estimate the difference between the two charge states ($\Delta B$) by computing
\begin{align} 
T_1^{\Delta B, \rm inel}(Q^2) &= \ffr{Q^2 2M_{\bar B} \beta^{\Delta B}/\alpha + Q^4C^{\Delta B}/(3\Lambda_{\beta}^2)^3}{(1+Q^2/(3\Lambda_{\beta}^2))^3} \, ,
\nn
\\ 
\label{T1_inel}
\end{align} 
where $M_{\bar B}$ is the average mass of the two charge states, $\beta^{\Delta B}$ is the difference between the magnetic polarizabilites between the two charge states and $\Lambda_{\beta}$ is a mass scale characterizing the Compton interaction. The factor $C^{\Delta B}$ describes the large-$Q^2$ behavior of the inelastic subtraction component:
\begin{align} 
  C^{\Delta B} &= C^{\Delta B}_{(0)}+C^{\Delta B}_{(2)},\label{CDelta}\\
  C^{\Delta B}_{(0)}&= -4M_{\bar B} \big(\mathcal{Q}_u^2 \fr{m_u}{\bar m} - \mathcal{Q}_d^2\fr{m_d}{\bar m} \big) \big(\sigma_u^{\bar B} - \sigma_d^{\bar B} \big) ,\\
  C^{\Delta B}_{(2)}&= 4M_{\Xi_{cc}}^2(\mathcal{Q}_u^2-\mathcal{Q}_d^2)\langle x\rangle^{\Xi_{cc}}_l.
\end{align} 
The light quark masses are denoted by $m_{u,d}$, and $\bar m$ represents the average.
The factor $\sigma_q^{\bar B}$ is the average sigma term for the quark
$q$ in the pair of $\Xi_{cc}$ charge states, e.g. $\sigma_u^{\bar B}=(\sigma_u^{\Xi_{cc}^{++}}+\sigma_u^{\Xi_{cc}^{+}})/2$.
We estimate
this based on the calculated baryon octet sigma terms in lattice QCD
analyses~\cite{1206.3156,Shanahan:2012wa}, which find $\sigma_u^N - \sigma_d^N =
-13(2)$ MeV, $\sigma_u^{\Sigma} - \sigma_d^{\Sigma} = -6(1)$ MeV and
$\sigma_u^{\Xi} - \sigma_d^{\Xi} = -3(1)$ MeV. These indicate that the
difference between the $u$ and $d$ contributions to the mass of baryon
decreases as heavier flavors are introduced. Thus, we estimate
$\sigma_u^{\Xi_{cc}} - \sigma_d^{\Xi_{cc}}= -1$ MeV. While we allow
for 100\% variation in this value, it has no appreciable impact on the final result.

The spin-2 contribution to the asymptotic behavior of the subtraction
function was overlooked in Ref.~\cite{Erben:2014hza}, and has been recently
discussed in Ref.~\cite{Hill:2016bjv}. The key point of relevance to
the Cottingham integral is that the spin-2 contribution in the
subtraction function exactly cancels against the corresponding term in
the inelastic part of the dispersion integral
\cite{Collins:1978hi}. The electromagnetic quark self energy on the
sigma term remains as the only term that contributes to the
logarithmic running of the complete self energy of the nucleon.
Even though the spin-2 contribution is almost an order of magnitude
larger than the spin-0 contribution, it's omission in
Ref.~\cite{Erben:2014hza} has negligible effect on the final results reported.

Just as the lack of knowledge of the magnetic polarizability $\beta^{\Delta B}$ and the mass scale $\Lambda_{\beta}$ dominated the uncertainty in the calculation of the mass splitting between the octet baryons in Ref.~\cite{Erben:2014hza}, so it dominates in the present calculation. The best constraint on the magnetic polarizabilities of the nucleon using chiral effective field theory and experimental data are given in Ref.~\cite{1409.3705} to be 
\begin{align} 
\beta^p &= 3.2(5) \times 10^{-4} {\rm fm}^3\\
\beta^n &= 3.7(15) \times 10^{-4} {\rm fm}^3\\
\Rightarrow \beta^{\Delta N} &= -0.5(16) \times 10^{-4} {\rm fm}^3 \, .
\end{align}
A recent lattice QCD analysis~\cite{1707.02765} of the magnetic polarizabilities of the proton shows that for non-physical heavy pion masses, the magnetic polarizability remains relatively constant. Based on this observation and a lack of results for the $\Xi_{cc}$ magnetic polarizabilities, we follow Ref.~\cite{Erben:2014hza} by taking the same value and uncertainty range as for the nucleon polarizability.  

To first order in $Q^2$, the inelastic $T_1^{B, \rm inel}$ amplitude term at low $Q^2$ is given by the magnetic polarizability. Motivating the equation~\ref{T1_inel}, we note that the next to leading order in $Q^2$ is determined by the polarizability form factor $F_{\beta}$, as given in the chiral perturbation theory analysis for the nucleon by Birse and McGovern~\cite{1206.3030}. There, the form factor is given as 
\begin{align} 
F_{\beta} &= 1 + \ffr{Q^2}{[\Lambda_{\beta}^N]^2} + \mathcal{O}(Q^4) \, ,
\end{align} 
where the expansion depends on the mass scale $\Lambda_{\beta}^N$ = 460(108) MeV \cite{1206.3030}. In the dispersion relation estimate of the octet baryons of Ref.~\cite{Erben:2014hza}, a heavier mass scale is used, based on the fact that the physics is governed by the interaction with the heavier $s$ quarks of the $\Sigma^{+/-}$ and $\Xi^{0/-}$ baryons. The mass scale used in that analysis is $\Lambda_{\beta}^{\Sigma,\Xi} = 0.7(3)$ GeV. Here, considering the massive $c$ quark contribution to the mass scale, we choose $\Lambda_{\beta}^{\Xi_{cc}}$  = 1.0(3) GeV. 

\begin{figure}
\begin{center}
\caption{(Color online) Subtraction inelastic integrand as function of $Q^2$ for $\Xi_{cc}$ and nucleon contributions to electromagnetic self-energy. In both cases of the $\Xi_{cc}$ (lavender) and the nucleon (dark blue) integrands, the shaded regions reflect the uncertainty in the integrands due to the uncertainty in the magnetic polarizabilty difference between the charge states, $\beta^{\Delta B}$.}
\includegraphics[width = 1.0\linewidth]{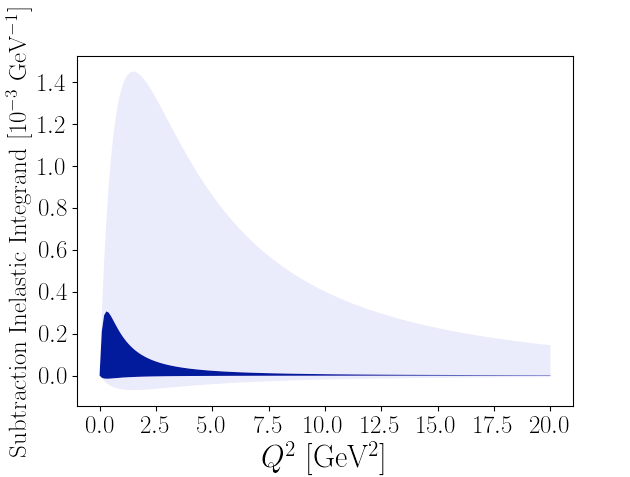}
\label{sub_inel_compare}
\end{center}
\end{figure}

Using the parameters above, we compute the elastic and inelastic subtraction terms summarized in Table~\ref{results_table} -- see Fig.~\ref{sub_inel_compare} to compare the subtraction inelastic integrand for the $\Xi_{cc}$ and the nucleon. The shaded regions represent the uncertainty in the difference in magnetic polarizability of the charge states. Note that the range of reasonable values for the subtraction inelastic contribution for the $\Xi_{cc}$ is much larger than that for the nucleon, allowing for a much greater contribution to the mass splitting between the $\Xi_{cc}^{++}$ and the $\Xi_{cc}^+$. 
%

\subsection{Counter Terms}
The final contributions in the dispersion analysis are the counter terms, which are given to account for the scale dependence determined by $\Lambda_0$. Following the analysis for the computation of the baryon octet charge symmetry violation of Ref.~\cite{Erben:2014hza}, we take the leading order contribution to be 
\begin{align} 
\delta M_{\Delta B}^{\rm ct} &= -\ffr{3\alpha}{16\pi M_{\bar B}} C^{\Delta B}_{(0)} \log\Big( \ffr{\Lambda_0^2}{\Lambda_1^2}\Big) \, ,
\end{align} 
where we follow Ref.~\cite{WalkerLoud:2012bg} in taking $\Lambda_1^2$ = 100 GeV$^2$. The large value of $M_{\bar B}$ and the small value of $C^{\Delta B}$, as obtained in Eq.~\ref{CDelta}, yield a value on the order of 10$^{-4}$ MeV for $\delta M_{\Xi_{cc}}^{\rm ct}$. As a result we include the result in Table~\ref{results_table} as $\approx$0.

\section{Total}
We have calculated a conservative estimate of the total electromagnetic symmetry breaking in the doubly charmed cascade baryons using a subtracted dispersion relation. In summary, we find 
\begin{align}
\delta M_{\Xi_{cc}^{++}}^{\gamma} - \delta M_{\Xi_{cc}^{+}}^{\gamma} = 8(9) {\rm \, MeV} \, . 
\end{align}
In Table~\ref{results_table}, we summarize each of the contributions to the total electromagnetic self energy, as given by the subtracted dispersion relation analysis (Eq.~\ref{contributions}). The greatest contributions come from the elastic and subtraction inelastic terms. As depicted in Fig.~\ref{elastic_compare}, the elastic term is greater for the $\Xi_{cc}$ than the nucleon. This is attributed to the small charge radii for the $c$ system, as well as a small magnetic moment of the $\Xi_{cc}^+$. Thus, the investigation of the quark-diquark model of the doubly charmed baryons is important to our understanding of the charge symmetry violation in these exotic particles. Similarly, we may compare the next largest term, the subtraction inelastic term, calculated for the $\Xi_{cc}$ to that of the nucleon (see Eq.~\ref{T1_inel}). The large contribution from this term can be attributed to the larger mass $M_B$ for the $\Xi_{cc}$ and the largeness of the mass scale associated with the magnetic polarizabilty, $\Lambda_{\beta}$ (see Fig.~\ref{sub_inel_compare}). 

In Table~\ref{uncertainties} we summarize the parameters and uncertainty ranges used in each of the elastic, inelastic, subtraction and counter terms. The error propagation due to each parameter is shown in the uncertainty yield column. From the table, we see that calculated mass splitting is relatively stable for the range of masses for the $\Xi_{cc}$. As described earlier, within the range 10 GeV$^2 < \Lambda_0^2 < $ 30 GeV$^2$, variation in the renormalization scale $\Lambda_0$ does not effect the result.
By far the largest source of uncertainty in the calculation is that associated with the magnetic polarizabilty and its mass scale. 
\begin{table*}[t]
\begin{center}
\caption{Decomposition of the electromagnetic contributions to the $\Xi_{cc}$ baryon mass splittings as defined in Eq. \ref{contributions}. We also list the results of the dispersive analysis of Ref.~\cite{Erben:2014hza} for the nucleon electromagnetic self-energy contributions for comparison. All masses given in MeV. }

\setlength{\tabcolsep}{12pt}
\begin{tabular}{|c|c c c c c|c |}
\hline 
Baryon & $\delta M^{\rm el}$ & $\delta M^{\rm inel}$ & $\delta M^{\rm sub, el}$ & $\delta M^{\rm sub, inel}$ & $\delta M^{\rm ct}$ & $\delta M^{\gamma}$ \\
\hline 
$\Xi_{cc}^{++} - \Xi_{cc}^{+}$ & 3.19(35) & 0.30(10) & 1.65(36) & 3(9) & $\approx$ 0 & 8(9) \\
\hline 
$p - n$ & 1.401(7) & 0.089(42) & -0.635(7) & 0.18(35) & 0.006 & 1.04(35)\\
\hline 
\end{tabular}
\label{results_table}
\end{center}
\end{table*}
\begin{table*}[t]
\begin{center}
\caption{Parameters and contributions to overall uncertainty for each term of Eq. \ref{contributions}.}
\renewcommand{\arraystretch}{1.5}
\begin{tabular}{|c|c|c|c|c|}
\hline 
Term & Parameter & Value & Contribution to Total & Reference\\
	&			&		& Uncertainty (MeV) & 		\\
\hline 
\hline
All & $\Lambda_0$ (GeV)& 20(10) & $\approx$ 0& \\
\hline 
Elastic \& & $M_B \equiv  M_{\bar{B}}$ (MeV) & 3519(100) & 0.02(el), 0.02(sub) &  \cite{hep-ex/0208014,hep-ex/0212029,selex3}\\
Subtraction Elastic & $\<r_{E, c}^2\>$ (fm$^2$) & 0.095(9)&0.04(el), 0.01(sub) & \cite{Can:2013tna} \\
& $\<r_{E, l}^2\>$ (fm$^2$) & 0.410(46)&0.16(el), -0.04(sub) & \cite{Can:2013tna} \\
& $\<r_{M, c}^2\>$ (fm$^2$) & 0.089(11) &0.03(el), 0.04(sub) & \cite{Can:2013tna} \\
& $ \<r_{M, l}^2\>$ (fm$^2$) & 0.612(115) & 0.13(el), 0.16(sub) &\cite{Can:2013tna} \\
& $\mu_{\Xi_{cc}^+}$ ($\mu_N$) & 0.75(0.3) &0.27(el), 0.316(sub) & \cite{Can:2013tna, Barik:1984tq, JuliaDiaz:2004vh, Kumar:2005ei, hep-ph/0602193, hep-ph/0610030, 1003.4338, 1209.2900,1707.02765}\\
& $\mu_{\Xi_{cc}^{++}}$ ($\mu_N$) & -0.10(0.10) & 0.01(el), 0.01(sub)&\cite{Can:2013tna, Barik:1984tq, JuliaDiaz:2004vh, Kumar:2005ei, hep-ph/0602193, hep-ph/0610030, 1003.4338, 1209.2900,1707.02765}\\
\hline 
Inelastic &  $M_B \equiv  M_{\bar{B}}$ (MeV) & 3519(100) & $\approx 0$ &  \cite{hep-ex/0208014,hep-ex/0212029,selex3}\\
& ${\<x\>^{\Xi_{cc}}_l}/{\<x\>^p_d}$ & 0.30(10)& 0.10 &\cite{Erben:2014hza}\\ 
\hline 
Subtraction Inelastic & $M_B \equiv  M_{\bar{B}}$ (MeV) & 3519(100) & $\approx 0$ &  \cite{hep-ex/0208014,hep-ex/0212029,selex3}\\
& $\beta^{\Delta B} \equiv \beta^{\Xi_{cc}^{++}} - \beta^{\Xi_{cc}^+}$ (fm$^3$) & -0.5(1.6) $\times 10^{-4}$ &8.5& \\ 
& $\Lambda_{\beta}$ (GeV) & 1.0(3) & 2.8&\cite{Erben:2014hza}\\
& $\sigma_u^{\Xi_{cc}} - \sigma_d^{\Xi_{cc}}$(MeV) & -1(1)& $\approx 0$&\\
\hline 
Counter & $M_B \equiv  M_{\bar{B}}$ (MeV) & 3519(100) & $\approx 0$ &  \cite{hep-ex/0208014,hep-ex/0212029,selex3}\\
& $\Lambda_1^2$ (GeV$^2$) & 100(100) &$\approx 0$& \cite{WalkerLoud:2012bg} \\ 
\hline 
\end{tabular}
\label{uncertainties}
\end{center}
\end{table*}
%

\section{Summary}
The large uncertainty range in our calculation of the electromagnetic self-energy is a result of the current uncertainty in the magnetic polarizability of baryons in general. However, within reasonable bounds on these parameters, we find that a relatively large CSV is possible for the $\Xi_{cc}$ system. The dispersion relation estimates of Ref.~\cite{Erben:2014hza} for the electromagnetic self-energy of the octet baryons give contributions to the CSV of about 1 MeV, which are comparable to lattice analyses as in Refs.~\cite{1406.4579, Borsanyi:2014jba, Borsanyi:2013lga}, as well as the analysis of Ref.~\cite{WalkerLoud:2012bg}. Thus, it is interesting to find the possibility of such a large electromagnetic self-energy for the doubly heavy cascade systems. 

It is of considerable experimental and theoretical interest to continue the investigation of the properties of doubly heavy baryons. However, we must face the sources of uncertainty in these calculations. In this respect, our current inability to experimentally access some variables such as magnetic polarizabilty of exotic baryons, means that in the near future we will need to look for future lattice simulations. As the magnetic polarizability information is the greatest source of uncertainty in the present calculation, we eagerly await results from lattice simulations to obtain a more precise understanding of the charge symmetry violation in the $\Xi_{cc}$ system, and for other exotic heavy baryons as well.

\section*{Acknowledgements}
KC thanks Robert Perry for useful discussions. This work was supported by the University of Adelaide and the Australian Research Council through the ARC Center of Excellence for Particle Physics at the Terascale and grants DP150103101 (AWT) and FT120100821, DP140103067 (RDY). 

\section*{References}

\bibliography{bib_cascades}

\begin{thebibliography}{10}
\expandafter\ifx\csname url\endcsname\relax
  \def\url#1{\texttt{#1}}\fi
\expandafter\ifx\csname urlprefix\endcsname\relax\def\urlprefix{URL }\fi
\expandafter\ifx\csname href\endcsname\relax
  \def\href#1#2{#2} \def\path#1{#1}\fi

\bibitem{Patrignani:2016xqp}
Review of particle physics, Chin.\ Phys.\ C 40~(10, 100001).

\bibitem{hep-ex/0208014}
M.~M. {\it et al.}~[SELEX~Collaboration], First observation of the doubly
  charmed baryon xi+(cc), Phys.\ Rev.\ Lett.\ 89~(112001).
\newblock \href {http://dx.doi.org/10.1103/PhysRevLett.89.112001}
  {\path{doi:10.1103/PhysRevLett.89.112001}}.

\bibitem{hep-ex/0212029}
M.~A.~M. {\it et al.}~[SELEX~Collaboration], First observation of doubly
  charmed baryons, Czech.\ J.\ Phys.\ 53~(B201).
\newblock \href {http://dx.doi.org/10.1103/PhysRevLett.89.112001}
  {\path{doi:10.1103/PhysRevLett.89.112001}}.

\bibitem{selex3}
The double charm baryon family at selex: An update, Fermilab Joint Experimental
  and Theoretical Physics Seminar.

\bibitem{Brodsky:2011zs}
C.~H. S.~J.~Brodsky, F.~K.~Guo, U.~G. Meissner, Isospin splittings of doubly
  heavy baryons, Phys.\ Lett.\ B 698~(251).
\newblock \href {http://dx.doi.org/10.1016/j.physletb.2011.03.014}
  {\path{doi:10.1016/j.physletb.2011.03.014}}.

\bibitem{Can:2013tna}
B.~I.~M.~O. K.~U.~Can, G.~Erkol, T.~T. Takahashi, Electromagnetic structure of
  charmed baryons in lattice qcd, JHEP 1405~(125).
\newblock \href {http://dx.doi.org/10.1007/JHEP05(2014)125}
  {\path{doi:10.1007/JHEP05(2014)125}}.

\bibitem{Beane:2006fk}
K.~O. S.~R.~Beane, M.~J. Savage, Strong-isospin violation in the neutron-proton
  mass difference from fully-dynamical lattice qcd and pqqcd, Nucl.\ Phys.\ B
  768~(38).
\newblock \href {http://dx.doi.org/10.1016/j.nuclphysb.2006.12.023}
  {\path{doi:10.1016/j.nuclphysb.2006.12.023}}.

\bibitem{Horsley:2012fw}
R.~H. {\it et al.}~[QCDSF, U.~Collaborations], sospin breaking in octet baryon
  mass splittings, Phys.\ Rev.\ D 86~(114511).
\newblock \href {http://dx.doi.org/10.1103/PhysRevD.86.114511}
  {\path{doi:10.1103/PhysRevD.86.114511}}.

\bibitem{Shanahan:2012wa}
A.~W.~T. P.~E.~Shanahan, R.~D. Young, Strong contribution to octet baryon mass
  splittings, Phys.\ Lett.\ B 718~(1148).
\newblock \href {http://dx.doi.org/10.1016/j.physletb.2012.11.072}
  {\path{doi:10.1016/j.physletb.2012.11.072}}.

\bibitem{Borsanyi:2014jba}
S.~B. {\it et al.}, Ab initio calculation of the neutron-proton mass
  difference, Science 347~(1452).
\newblock \href {http://dx.doi.org/10.1126/science.1257050}
  {\path{doi:10.1126/science.1257050}}.

\bibitem{Horsley:2015eaa}
R.~H. {\it et al.}, Isospin splittings of meson and baryon masses from
  three-flavor lattice qcd + qed, J.\ Phys.\ G 43~(10, 10LT02).
\newblock \href {http://dx.doi.org/10.1088/0954-3899/43/10/10LT02}
  {\path{doi:10.1088/0954-3899/43/10/10LT02}}.

\bibitem{Cottingham:1963zz}
W.~N. Cottingham, The neutron proton mass difference and electron scattering
  experiments, Annals Phys.\ 25~(424).
\newblock \href {http://dx.doi.org/10.1016/0003-4916(63)90023-X}
  {\path{doi:10.1016/0003-4916(63)90023-X}}.

\bibitem{Gasser:1974wd}
J.~Gasser, H.~Leutwyler, Implications of scaling for the proton - neutron mass
  - difference, Nucl.\ Phys.\ B 94~(269).
\newblock \href {http://dx.doi.org/10.1016/0550-3213(75)90493-9}
  {\path{doi:10.1016/0550-3213(75)90493-9}}.

\bibitem{Collins:1978hi}
J.~C. Collins, Renormalization of the cottingham formula, Nucl.\ Phys.\ B
  149~(90).
\newblock \href {http://dx.doi.org/10.1016/j.nuclphysb.2016.12.017,
  10.1016/0550-3213(79)90158-5} {\path{doi:10.1016/j.nuclphysb.2016.12.017,
  10.1016/0550-3213(79)90158-5}}.

\bibitem{WalkerLoud:2012bg}
C.~E.~C. A.~Walker-Loud, G.~A. Miller, The electromagnetic self-energy
  contribution to $m_p - m_n$ and the isovector nucleon magnetic
  polarizability, Phys.\ Rev.\ Lett.\ 108~(232301).
\newblock \href {http://dx.doi.org/10.1103/PhysRevLett.108.232301}
  {\path{doi:10.1103/PhysRevLett.108.232301}}.

\bibitem{Gasser:2015dwa}
H.~L. J.~Gasser, M.~Hoferichter, A.~Rusetsky, Cottingham formula and nucleon
  polarisabilities, Eur.\ Phys.\ J.\ C 75~(8, 375).
\newblock \href {http://dx.doi.org/10.1140/epjc/s10052-015-3580-9}
  {\path{doi:10.1140/epjc/s10052-015-3580-9}}.

\bibitem{Erben:2014hza}
A.~W.~T. F.~B.~Erben, P.~E.~Shanahan, R.~D. Young, Dispersive estimate of the
  electromagnetic charge symmetry violation in the octet baryon masses, Phys.\
  Rev.\ C 90~(6, 065205).
\newblock \href {http://dx.doi.org/10.1103/PhysRevC.90.065205}
  {\path{doi:10.1103/PhysRevC.90.065205}}.

\bibitem{Aaij:2017ueg}
R.~A. {\it et al.}~[LHCb~Collaboration], Observation of the doubly charmed
  baryon $\xi_{cc}^{++}$, Phys.\ Rev.\ Lett.\ 119~(11, 112001).
\newblock \href {http://dx.doi.org/10.1103/PhysRevLett.119.112001}
  {\path{doi:10.1103/PhysRevLett.119.112001}}.

\bibitem{Barik:1984tq}
N.~Barik, M.~Das, Magnetic moments of confined quarks and baryons in an
  independent quark model based on dirac equation with power law potential,
  Phys.\ Rev.\ D 28~(2823).
\newblock \href {http://dx.doi.org/10.1103/PhysRevD.28.2823}
  {\path{doi:10.1103/PhysRevD.28.2823}}.

\bibitem{JuliaDiaz:2004vh}
B.~Julia-Diaz, D.~O. Riska, Baryon magnetic moments in relativistic quark
  models, Nucl.\ Phys.\ A 739~(69).
\newblock \href {http://dx.doi.org/10.1016/j.nuclphysa.2004.03.078}
  {\path{doi:10.1016/j.nuclphysa.2004.03.078}}.

\bibitem{Kumar:2005ei}
R.~D. S.~Kumar, R.~C. Verma, Magnetic moments of charm baryons using effective
  mass and screened charge of quarks, J.\ Phys.\ G 31~(2, 141).
\newblock \href {http://dx.doi.org/10.1088/0954-3899/31/2/006}
  {\path{doi:10.1088/0954-3899/31/2/006}}.

\bibitem{hep-ph/0602193}
M.~A.~I.~J.~G.~K. V.~E.~L. D.~N. A.~Faessler, T.~Gutsche, K.~Pumsa-ard,
  Magnetic moments of heavy baryons in the relativistic three-quark model,
  Phys.\ Rev.\ D 73~(094013).
\newblock \href {http://dx.doi.org/10.1103/PhysRevD.73.094013}
  {\path{doi:10.1103/PhysRevD.73.094013}}.

\bibitem{hep-ph/0610030}
J.~N. C.~Albertus, E.~Hernandez, J.~M. Verde-Velasco, Static properties and
  semileptonic decays of doubly heavy baryons in a nonrelativistic quark model,
  Eur.\ Phys.\ J.\ A 32~(183).
\newblock \href {http://dx.doi.org/10.1140/epja/i2007-10364-y,
  10.1140/epja/i2008-10547-0} {\path{doi:10.1140/epja/i2007-10364-y,
  10.1140/epja/i2008-10547-0}}.

\bibitem{1003.4338}
P.~K.~C. N.~Sharma, H.~Dahiya, M.~Gupta, Spin 1/2$^+$, spin 3/2$^+$ and
  transition magnetic moments of low lying and charmed baryons, Phys.\ Rev.\ D
  81~(073001).
\newblock \href {http://dx.doi.org/10.1103/PhysRevD.81.073001}
  {\path{doi:10.1103/PhysRevD.81.073001}}.

\bibitem{1209.2900}
A.~Bernotas, V.~Simonis, {Magnetic moments of heavy baryons in the bag model
  reexamined }\href {http://arxiv.org/abs/1209.2900} {\path{arXiv:1209.2900}}.

\bibitem{1707.02765}
Z.~W.~L. H.~S.~Li, L.~Meng, S.~L. Zhu, Magnetic moments of the doubly charmed
  and bottom baryons, Phys.\ Rev.\ D 96~(7, 076011).
\newblock \href {http://dx.doi.org/10.1103/PhysRevD.96.076011}
  {\path{doi:10.1103/PhysRevD.96.076011}}.

\bibitem{Kelly:2004hm}
J.~J. Kelly, Simple parametrization of nucleon form factors, Phys.\ Rev.\ C
  70~(068202).
\newblock \href {http://dx.doi.org/10.1103/PhysRevC.70.068202}
  {\path{doi:10.1103/PhysRevC.70.068202}}.

\bibitem{Londergan:2009kj}
J.~C.~P. J.~T.~Londergan, A.~W. Thomas, Charge symmetry at the partonic level,
  Rev.\ Mod.\ Phys.\ 82~(2009).
\newblock \href {http://dx.doi.org/10.1103/RevModPhys.82.2009}
  {\path{doi:10.1103/RevModPhys.82.2009}}.

\bibitem{Gershtein:2000nx}
A.~K.~L. S.~S.~Gershtein, V.~V.~Kiselev, A.~I. Onishchenko, Spectroscopy of
  doubly heavy baryons, Phys.\ Rev.\ D 62~(054021).
\newblock \href {http://dx.doi.org/10.1103/PhysRevD.62.054021}
  {\path{doi:10.1103/PhysRevD.62.054021}}.

\bibitem{1206.3156}
R.~H. {\it et al.}~[QCDSF, U.~Collaborations], Isospin breaking in octet baryon
  mass splittings, Phys.\ Rev.\ D 86~(114511).
\newblock \href {http://dx.doi.org/10.1103/PhysRevD.86.114511}
  {\path{doi:10.1103/PhysRevD.86.114511}}.

\bibitem{Hill:2016bjv}
R.~J. Hill, G.~Paz, Nucleon spin-averaged forward virtual compton tensor at
  large $q^2$, Phys.\ Rev.\ D 95~(9, 094017).
\newblock \href {http://dx.doi.org/10.1103/PhysRevD.95.094017}
  {\path{doi:10.1103/PhysRevD.95.094017}}.

\bibitem{1409.3705}
L.~S.~M. {\it et al.}~[COMPTON@MAX-lab Collaboration], Measurement of compton
  scattering from the deuteron and an improved extraction of the neutron
  electromagnetic polarizabilities, Phys.\ Rev.\ Lett.\ 113~(26, 262506).
\newblock \href {http://dx.doi.org/10.1103/PhysRevLett.113.262506}
  {\path{doi:10.1103/PhysRevLett.113.262506}}.

\bibitem{1206.3030}
M.~C. Birse, J.~A. McGovern, Proton polarisability contribution to the lamb
  shift in muonic hydrogen at fourth order in chiral perturbation theory, Eur.\
  Phys.\ J.\ A 48~(120).
\newblock \href {http://dx.doi.org/10.1140/epja/i2012-12120-8}
  {\path{doi:10.1140/epja/i2012-12120-8}}.

\bibitem{1406.4579}
X.~G.~W. A.~W.~Thomas, R.~D. Young, Electromagnetic contribution to the
  proton-neutron mass splitting, Phys.\ Rev.\ C 91~(1, 015209).
\newblock \href {http://dx.doi.org/10.1103/PhysRevC.91.015209}
  {\path{doi:10.1103/PhysRevC.91.015209}}.

\bibitem{Borsanyi:2013lga}
S.~B. {\it et al.}~[Budapest-Marseille-Wuppertal~Collaboration], Isospin
  splittings in the light baryon octet from lattice qcd and qed, Phys.\ Rev.\
  Lett.\ 111~(25, 252001).
\newblock \href {http://dx.doi.org/10.1103/PhysRevLett.111.252001}
  {\path{doi:10.1103/PhysRevLett.111.252001}}.

\end{thebibliography}

\end{document}